\documentclass[letterpaper, 11pt]{article}

\usepackage{geometry}
\usepackage[utf8]{inputenc}
\usepackage{xcolor}
\usepackage{amsmath,amsthm,amsfonts,amssymb}
\usepackage{tikz}
\usepackage{enumitem}
\usepackage{tabularx}
\usepackage{hyperref}
\usepackage[sort, numbers, compress]{natbib}
\usepackage[normalem]{ulem}
\usepackage[ruled,vlined]{algorithm2e}

\usepackage{cite}
\usepackage{algorithmic}
\usepackage{graphicx}
\usepackage{textcomp}

\tikzstyle{dot}=[circle,fill,inner sep=1.5pt]

\newtheorem{lemma}{\bfseries Lemma}
\newtheorem{claim}{\bfseries Claim}

\providecommand{\norm}[1]{\ensuremath{\left\lVert#1\right\rVert }}
\providecommand{\mnorm}[1]{\ensuremath{\left\lvert#1\right\rvert}}

\providecommand{\comment}[1]{}
\providecommand{\nitin}[1]{}

\def\R{\mathbb{R}}
\def\H{\mathcal{H}}

\def\W{\mathcal{W}}

\begin{document}

\title{Byzantine Fault-Tolerant Min-Max Optimization\thanks{This version adds necessary citations and fixings in the proofs in Section~\ref{sec:algo-distributed}.}}
\author{Shuo Liu \thanks{Georgetown University. Email: {\tt sl1539@georgetown.edu}.} \hspace{0.5in} Nitin H. Vaidya \thanks{Georgetown University. Email: {\tt nitin.vaidya@georgetown.edu}.}
}
\author{Shuo Liu \hspace{0.5in} Nitin H. Vaidya\\\small{Georgetown University}\\\small{\texttt{\{sl1539, nitin.vaidya\}@georgetown.edu}}}
\date{}

\maketitle

\begin{abstract}
    In this paper, we consider a min-max optimization problem under adversarial manipulation, where there are $n$ cost functions, up to $f$ of which may be replaced by arbitrary \textit{faulty} functions by an adversary. The goal is to minimize the maximum cost over $x$ among the $n$ functions despite the faulty functions. The problem formulation could naturally extend to Byzantine fault-tolerant \textit{distributed} min-max optimization. 
    
    We present a simple algorithm for Byzantine min-max optimization, and provide bounds on the output of the algorithm. We also present an approximate algorithm for this problem. We then extend the problem to a distributed setting and present a distributed algorithm.
    To the best of our knowledge, we are the first to consider this problem.
\end{abstract}

\section{Introduction}
\label{sec:intro}


Suppose there is a set of cost functions $Q_i(x)$, $i\in\{1,...,n\}$ and $x\in \W$, where $\W\subset\R^d$ is a region that the problem is defined on. 
The goal of the (fault-free) min-max problem considered here is to compute an element in:
\begin{align}
    \arg\min_{x\in\W}\max_{i\in[n]}Q_i(x),
    \label{eqn:min-max-fault-free}
\end{align}
where $[n]$ denotes the set $\{1,...,n\}$. Intuitively speaking, we seek a minimum of the worst-case cost among all $n$ functions. In the distributed version of the problem, there are $n$ agents in the system, and each agent $i$ possesses one local cost function $Q_i(x)$; the agents collaboratively compute an optimal $x$ as specified in \eqref{eqn:min-max-fault-free}. As a simple example, consider a group of $n$ people deciding a place to meet. Suppose that the cost function of the $i$-th person is the time it takes to travel from their current place to the meeting place. The goal is to minimize the overall time it would take for all people to arrive at the meeting place, i.e., the longest time it takes for anyone to arrive. Previous work has studied this (fault-free) problem and proposed a distributed algorithm in a decentralized network \citep{srivastava2013distributed,srivastava2011distributed}.

In this paper, we consider such min-max optimization, but with up to $f$ of the cost functions being replaced by arbitrary faulty functions by an adversary. In a distributed setting, this corresponds to up to $f$ agents being Byzantine faulty -- a Byzantine faulty agent can behave arbitrarily \citep{lamport2019byzantine}. We allow a faulty function to be arbitrarily ``bad''. Such faulty functions can render the result of \eqref{eqn:min-max-fault-free} meaningless. For instance, recall our example of the meeting above. It should be easy to see that an adversary may provide a fabricated (faulty) cost function such that it can completely control the outcome of \eqref{eqn:min-max-fault-free}. In other words, in our example, the adversary can effectively force everyone to meet at a location of its choice, by claiming one of the cost functions to always be the maximum, and minimize at its desired location. The question then is what should be done so that the adversary's impact on the choice of the meeting place is constrained. We approach this question by defining a modified, \textit{reasonable} goal for the fault-tolerant min-max optimization. In particular, our ``ideal'' goal for Byzantine fault-tolerant min-max optimization is to compute an element in: 
\begin{align}
    \arg\min_{x\in\W}\max_{i\in\H}Q_i(x),
    \label{eqn:min-max}
\end{align}
where $\H\subseteq[n]$ is the set of indices for the non-faulty functions (or, non-faulty agents, in a distributed setting). It is worth noting that we do not know, a priori, which functions are non-faulty.

It can be impractical to correctly identify the faulty functions since a clever adversary may hide its malicious behavior behind seemingly normal behaviors.
Also, the non-faulty functions may be independent of each other. Thus, in general, it is not easy or even possible to achieve the ideal goal in \eqref{eqn:min-max}. However, we can design algorithms that attempt to \emph{approximate} the solution of \eqref{eqn:min-max} in some desirable manner. 

\paragraph{Contributions of this report}
We introduce the problem of Byzantine fault-tolerant min-max optimization. We provide a centralized algorithm and obtain bounds on their behavior while tolerating up to $f$ faulty cost functions in Section~\ref{sec:algo_1}. We also present an algorithm that approximates the behavior of our first algorithm in Section~\ref{sec:algo-2}. We then provide a distributed algorithm and analyze its convergence in Section~\ref{sec:algo-distributed}.
This paper opens up a new line of research, with much work needed to develop efficient \emph{distributed} algorithms.



\section{Related work}
\label{sec:related}

There is significant work on a different formulation of fault-tolerant distributed optimization. In particular, that line of work
focuses on finding minimum of the {\em sum} of the different agents' cost functions \citep{su2016fault, liu2021approximate, chen2017distributed, blanchard2017machine, yin2018byzantine}. That is, the goal there is to find (or approximate) 
\[
\arg \min_{x} \sum_{i\in \H} Q_i(x)
\]
where $\H$ is the set of fault-free agents, and $Q_i(x)$ is the $i$-th agent's cost function. The problem formulation considered in our work replaces the summation above with a maximum.

Closer to our problem formulation, previous work in \citep{adibi2022distributed, lin2020near, daskalakis2018limit} has explored the following problem (in the absence of faults):
\begin{align}
    \min_{x\in\mathcal{X}}\max_{y\in\mathcal{Y}}Q(x,y)
    \label{eqn:min-max-alt}
\end{align}
where $Q(x,y)$ is a function over two vector-valued variables $x\in\mathcal{X}$ and $y\in\mathcal{Y}$. This problem has many applications, including game theory \citep{von1947theory}, online learning \citep{cesa2006prediction}, robust optimization \citep{ben2009robust}, and reinforcement learning \citep{dai2017learning}. \citet{yang2018cooperative} explored the distributed adaptation of this problem.

\citet{adibi2022distributed} studied the distributed min-max problem \eqref{eqn:min-max-alt} under the assumption that there are Byzantine agents in the system. It is assumed that each agent is able to compute stochastic gradients of the cost function $Q(x,y)$. In essences, all the agents have identical cost functions. The reason to use multiple agents in this setting is for speed-up, and Byzantine fault-tolerance is required when some of the agents may supply incorrect gradients. The problem formulation considered in our work differs from \citep{adibi2022distributed} in that we allow the different agents to have independent cost functions.

\citet{srivastava2013distributed} studied a distributed algorithm for the fault-free version of the min-max optimization problem \eqref{eqn:min-max-fault-free}. However, our work differs in that we allow for some of the cost functions to be faulty or tampered.
%
%
Previous work has also considered the problem of minimizing the maximum loss \citep{shalev2016minimizing, zhou2018minimax}, however, the problem formulation of this line of work is different from ours.

\citet{stolz_et_al:LIPIcs:2016:6591} considered the problem of computing median of the inputs of $n$ nodes, in the presence of faulty nodes. 
The similarity between the min-max problem and median is that both problems relate to rank -- we consider max (rank 1) over \emph{functions}, whereas \citep{stolz_et_al:LIPIcs:2016:6591} considers median (rank $n/2$) over \emph{scalars}. 

DIRECT is an algorithm proposed in \citep{jones1993lipschitzian} that solves optimization problems on Lipschitzian but not necessarily convex functions. The algorithm samples the function values of an everywhere dense point set, such that with Lipschitzness of the function, the minimum sampled function value can be arbitrarily close to the actual minimum asymptotically. As we will discuss later in Sections~\ref{sec:gradient-method} and~\ref{sec:algo-distributed}, in Byzantine min-max problems, we need to find the minimum of a non-convex function, where gradient descent-based algorithms are not suitable, but algorithms like DIRECT would be. 
\section{A Byzantine fault-tolerant algorithm and its performance bounds}
\label{sec:algo_1}
Recall that there are $n$ cost functions $Q_i(x)$'s, where up to $f$ functions may be replaced by faulty functions by an adversary. 
It can be shown that if the algorithm do not have any prior knowledge on whether a cost function is faulty or not, the number of agents $n\geq 2f+1$.
\comment{Nitin's comment: need to offer some proof of this - what does "not feasible" mean?\\\textcolor{black}{TODO: The original sentence is removed. It was about $n\geq 2f+1$. Discussion of this will be added later, together with the necessity of redundancy.}} 
We assume that a minimum exists for each non-faulty function $Q_i(x)$. Let $\H$ denote the set of indices of the non-faulty functions.
Our goal is to solve the  problem
\begin{align*}
    \arg\min_x\max_{i\in\H} Q_i(x).
\end{align*}

In the presence of faulty functions, in general, the above problem cannot be solved exactly (note that which functions are faulty is not known a priori). We will present bounds on how well we can approximate the ideal solution of the above problem using the algorithm presented below. 



\noindent\textbf{Byzantine Fault-tolerant algorithm.} The algorithm
 outputs the minimum of the following function:
\begin{align}
    g_{f+1}(x) = \underset{i\in[n]}{\mathrm{rank}_{f+1}}(Q_i(x)),
    \label{eqn:hf}
\end{align}
where $\mathrm{rank}_k(\cdot)$ is the $k$-th largest value among a group of values. In particular, the algorithm outputs $\widehat{x}$ such that
$$\widehat{x}\in \arg\min_x g_{f+1}(x).$$

\noindent\textbf{Bounds for the fault-tolerant algorithm.}
Let us denote
$$\widehat{v} = g_{f+1}(\widehat{x}) = \underset{i\in[n]}{\mathrm{rank}_{f+1}}(Q_i(\widehat{x}))=\min_x \underset{i\in[n]}{\mathrm{rank}_{f+1}}(Q_i(x)).$$
The rank $r$ value among function values indexed by a set of integers $S\subseteq[n]$ 
is defined as follows: sort the values $Q_i(x)$ at some $x\in\R$ for a set $S$, $i\in S$, in a decreasing order, breaking ties by sorting indices $i$ in an increasing order. Rank 1 is the largest value, and Rank 2 is the next largest value, and so on. In general, rank $r$ is the $r$-th largest value in the sorted order of the $n$ values. Note that, in the case of a tie, rank $r$ and rank $r+1$ values may be identical.
%
The main result of this section is stated in Claim \ref{claim:1} below.
\begin{claim}
\label{claim:1} Recall that $\widehat{v} = g_{f+1}(\widehat{x}) = \underset{i\in[n]}{\mathrm{rank}_{f+1}}(Q_i(\widehat{x}))$. Then,
    \begin{align}
        \min_{x\in\R}\underset{i\in\H}{\mathrm{rank}_1}(Q_i(x)) \geq \widehat{v}
        \geq \underset{i\in\H}{\mathrm{rank}_{f+1}}(Q_i(\widehat{x}))
        \geq\min_{x\in\R} \underset{i\in\H}{\mathrm{rank}_{f+1}}(Q_i(x)).
        \label{eqn:claim-1-bound}
    \end{align}
\end{claim}

\begin{proof}
Let us define\footnote{Note that $\min$ denotes $\min_x$ even if the argument $x$ is not stated explicitly.}
    \begin{align}
        v_a & = \min\underset{i\in\H}{\mathrm{rank}_1}(Q_i(x)), \\
        x_a &\in \arg\min\underset{i\in\H}{\mathrm{rank}_1}(Q_i(x)). 
    \end{align}

    We first show that $v_a \geq \widehat{v}$.
    There are at most $f$ faulty functions, so the smallest rank\footnote{Recall that, in case of a tie, the same value will appear at different ranks.} of value $v_a$ at $x_a$  among all $n$ functions is at most $f+1$; let $r$ be the smallest rank of $v_a$, with $r\leq f+1$. We have
    \begin{align*}
        v_a&=\underset{i\in[n]}{\mathrm{rank}_{r}}(Q_i(x_a))\geq\min\underset{i\in[n]}{\mathrm{rank}_{r}}(Q_i(x)).
    \end{align*}
    Let $x'\in\arg\min\underset{i\in[n]}{\mathrm{rank}_{r}}(Q_i(x))$. We further have
    \begin{align*}
        v_a\geq&~\min\underset{i\in[n]}{\mathrm{rank}_{r}}(Q_i(x))=\underset{i\in[n]}{\mathrm{rank}_{r}}(Q_i(x')) \tag{from above} \\
        ~~\geq&~\underset{i\in[n]}{\mathrm{rank}_{f+1}}(Q_i(x')) \tag{because $r\leq f+1$} \\
        ~~\geq&~\min\underset{i\in[n]}{\mathrm{rank}_{f+1}}(Q_i(x))
        ~=~\widehat{v}.
    \end{align*}
    Thus, $v_a\geq \widehat{v}$. \\
    
    Now,
        $\widehat{v}=\underset{i\in[n]}{\mathrm{rank}_{f+1}}(Q_i(\widehat{x})) \geq \underset{i\in\H}{\mathrm{rank}_{f+1}}(Q_i(\widehat{x})).$
    Thus,
    \begin{align*}
        \widehat{v}& \geq \underset{i\in\H}{\mathrm{rank}_{f+1}}(Q_i(\widehat{x})) \geq \min\underset{i\in\H}{\mathrm{rank}_{f+1}}(Q_i(x)).
    \end{align*}
    
    The above inequality, and the result $v_a\geq \widehat{v}$ shown earlier, together prove the claim.
\end{proof}

\noindent \textbf{Observation 1.}
The above claim provides upper and lower bounds on $\underset{i\in\H}{\mathrm{rank}_{f+1}}(Q_i(\widehat{x}))$, as a measure of the ``accuracy'' of the output $\widehat{x}$ of our algorithm. Specifically,
$$\min_{x\in\R}\underset{i\in\H}{\mathrm{rank}_1}(Q_i(x))
        ~\geq ~ \underset{i\in\H}{\mathrm{rank}_{f+1}}(Q_i(\widehat{x}))  ~\geq~\min_{x\in\R} \underset{i\in\H}{\mathrm{rank}_{f+1}}(Q_i(x)).$$

The first inequality above implies that at $\widehat{x}$, all-but-$f$ of the non-faulty functions are guaranteed to have value upper bounded by the $\min_{x\in\R}\underset{i\in\H}{\mathrm{rank}_1}(Q_i(x))$. In fact, Claim \ref{claim:1} also upper bounds all-but-$f$ non-faulty functions at by $\widehat{x}$ by $\widehat{v}$. \\

\noindent \textbf{Observation 2.}
The bounds in Claim~\ref{claim:1} are tight in a certain sense. Specifically, the following two results can be shown. 

\begin{itemize}
    \item It is possible to find groups of functions such that $\widehat{v}$ in \eqref{eqn:claim-1-bound} equals the upper bound $\min_{x\in\R}\underset{i\in\H}{\mathrm{rank}_1}(Q_i(x))$ and lower bound $\min_{x\in\R}\underset{i\in\H}{\mathrm{rank}_{f+1}}(Q_i(x))$.
    \begin{proof}
        Consider the case where the function value of a faulty cost function is always larger than all non-faulty cost functions, i.e., for every $j$ indexing faulty functions and every $i$ indexing non-faulty functions, $Q_j(x)>Q_i(x)$ for all $x\in\R$. In this case, we have
            $\underset{i\in[n]}{\mathrm{rank}_{f+1}}(Q_i(x)) = \underset{i\in\H}{\mathrm{rank}_1}(Q_i(x))$
        for all $x\in\R$. Therefore,
            $\min g_{f+1}(x) = \min\underset{i\in[n]}{\mathrm{rank}_{f+1}}(Q_i(x)) = \min\underset{i\in\H}{\mathrm{rank}_1}(Q_i(x))$
        and $\widehat{v}$ equals the upper-bound in \eqref{eqn:claim-1-bound}. 
        
        Consider the case where the function value of a faulty cost function is always smaller than all non-faulty cost functions, i.e., for every $j$ indexing faulty functions and every $i$ indexing non-faulty functions, $Q_j(x)<Q_i(x)$ for all $x\in\R$. Note that $n\geq2f+1$, so $\mnorm{\H}\geq f+1$. In this case, we have
            $\underset{i\in[n]}{\mathrm{rank}_{f+1}}(Q_i(x)) = \underset{i\in\H}{\mathrm{rank}_{f+1}}(Q_i(x))$
        for all $x\in\R$. Therefore, 
            $\min g_{f+1}(x) = \min\underset{i\in[n]}{\mathrm{rank}_{f+1}}(Q_i(x)) = \min\underset{i\in\H}{\mathrm{rank}_{f+1}}(Q_i(x))$.
        In this case, $\widehat{v}$ equals
         the lower-bound in \eqref{eqn:claim-1-bound}.
    \end{proof}
    \item Suppose that there exists
    an algorithm that produces outputs $x^*$ and $v^*$ and provides the following guarantee: for some $r<f+1$,
    \begin{eqnarray}
    \label{eq:r}
    v^*\geq\underset{i\in\H}{\mathrm{rank}_{r}}(Q_i(x^*)),
    \end{eqnarray}
    then, for any positive constant $V$, there exists an execution (i.e., non-faulty and faulty functions) for which $v^* > \min_{x\in\R}\underset{i\in\H}{\mathrm{rank}_1}(Q_i(x)) + V.$
    
    In other words, at output $x^*$, if the algorithm attempts to provide an upper bound $v^*$ on more non-faulty functions than that provided by
    Claim \ref{claim:1}, then the upper bound $v^*$ can be arbitrarily bad
    compared to $\min_{x\in\R}\underset{i\in\H}{\mathrm{rank}_1}(Q_i(x))$. 
    \begin{proof}
            For any constant $V>0$, let us construct such an execution now. Consider a set of $n$ functions such that the following is true:
    \begin{eqnarray}
    \label{eq:assume:r}
    Q_j(x) > \max_{f+1\leq i\leq n}Q_i(x) + V, \mbox{~for~} j\leq f \textrm{ and all }x.
    \end{eqnarray}
    
    In execution E1, suppose that functions 1 through $n-f$ are non-faulty, and functions $n-f+1$ through $n$ are faulty.
    Thus, in execution E1, $\H=\{1,\cdots,n-f\}$.
    Suppose that a given deterministic algorithm produces output $x^*$
    and $v^*$ such that \eqref{eq:r} holds true. Since $r<f+1$, assumption
    \eqref{eq:assume:r} implies that 
    \begin{eqnarray}
    \label{eq:V}
    v^* > \max_{f+1\leq i\leq n}Q_i(x^*) + V.
    \end{eqnarray}
    Now consider execution E2 with the same set of $n$ functions as in execution E1. In execution E2, suppose that
    functions 1 through $f$ are faulty, and functions $f+1$ through $n$ are non-faulty. Thus, in execution E2, $\H=\{f+1,\cdots,n\}$.
    Since the algorithm is presented with the same set of
    $n$ functions as in execution E1, in execution E2 also, it will output same $x^*$ and $v^*$ as in execution E1. Now,
    \begin{align*}
        v^* &>  \max_{f+1\leq i\leq n}Q_i(x^*) + V ~~~\mbox{due to \eqref{eq:V}} \\
            &= \underset{i\in\H}{\mathrm{rank}_{1}}Q_i(x^*) + V 
            \\
            \Rightarrow~ v^* &>\min_x \underset{i\in\H}{\mathrm{rank}_{1}}Q_i(x) + V.
    \end{align*}
    Hence, the proof.
    \end{proof}
\end{itemize}
\comment{With slight re-wording, the second proof above also applies to non-deterministic algorithms. We can discuss later.}

\section{An approximate Byzantine fault-tolerant algorithm}
\label{sec:algo-2}


The above algorithm uses a black box to determine the minimum of $g_{f+1}(x)$. In this section, we present an algorithm that only attempts to approximately estimate the solution of the algorithm presented above, in exchange for less computation and more flexibility.
We make two assumptions in the remainder of this section: (i) we assume that the cost functions are non-negative, and (ii) we consider a constrained version of the problem, with the assumption that the argument $x$ is constrained to a hypercube $X$. \\

Recall that $g_{f+1}(x) = \underset{i\in[n]}{\mathrm{rank}_{f+1}}(Q_i(x))$.
Let us also define
\begin{align}
    h_1(x) & = \underset{i\in\H}{\mathrm{rank}_1}(Q_i(x)),\textrm{ and } \\
    h_{f+1}(x) & = \underset{i\in\H}{\mathrm{rank}_{f+1}}(Q_i(x)).
\end{align}
Then $h_{f+1}(x)\leq g_{f+1}(x)\leq h_1(x).$ \\

Let us assume that the cost function $Q_i(x)$ of any non-faulty agent $i$ is $L$-Lipschitz continuous in $X$. Then,
for any $x_1,x_2\in X$,
$\mnorm{Q_i(x_1)-Q_i(x_2)} \leq L \norm{x_1-x_2}, ~\forall i\in\H$.
We do not make any assumptions about the
faulty functions (i.e., $Q_j(x),~j\not\in\H$).
Recall that we have assumed that the constraint set $X$ is a hypercube.


\noindent\textbf{Approximate Fault-Tolerant Algorithm.}
\begin{itemize}
    \item Partition $X$ into hypercubes denoted as $S_j$, with the property described next. Let $C_j$ denote the center of hypercube $S_j$ and let $d_j$ denote its diameter. Then, for a specified constant parameter $\epsilon$ (which we will call the {\em approximation constant}), the hypercubes $S_j$'s are chosen such that,
    \begin{align}
    g_{f+1}(C_j) - L d_j \geq (1-\epsilon)\, \min_i g_{f+1}(C_i),~~~~ \forall j \label{eq_d}
\end{align}
where the minimum is over all the hypercubes in the subdivision of $X$. 

The DIRECT algorithm \citet{jones2021direct} results in a subdivision that is similar to that performed by the above algorithm. 

\item Let $k=\arg\min_i g_{f+1}(C_i)$. Let $\widehat{x}=C_k$. The algorithm outputs $\widehat{x}$. 
Note that $\min_i g_{f+1}(C_i)=g_{f+1}(C_k)$.
\end{itemize}

\begin{claim}
    At the output $\widehat{x}$ of the approximate algorithm above, at least $|\H|-f$ of the non-faulty cost functions are upper bounded by $\left(\frac{1}{1-\epsilon}\right) \, \min_{x\in X} h_1(x)$.
    \label{claim:2}
\end{claim}
\begin{proof}
    We first consider a lemma that will be used later.
    \begin{lemma}
    \label{lemma:Lipschitz}
    Given that function $Q_i(x)$ for non-faulty agent $i$ is Lipschitz continuous in $X$ with parameter $L$, $h_1(x)$ is also Lipschitz continuous in $X$ with parameter $L$.
    \end{lemma}
    \begin{proof}[Proof of Lemma~\ref{lemma:Lipschitz}]
    Consider any two points $x_1,x_2\in X$. Without loss of generality, suppose that $h_1(x_1)\geq h_1(x_2).$
    Recall that $h_1(x) = \underset{i\in\H}{\mathrm{rank}_1}(Q_i(x)).$
    Let $q = \arg\max_{i\in\H} Q_i(x_1)$ and $r = \arg\max_{i\in\H} Q_i(x_2)$.
    Then, $h_1(x_1)=Q_q(x_1)$ and $h_1(x_2)=Q_r(x_2)$. 
    
    Now,
    $Q_q(x_2)\leq Q_r(x_2)$ because of the definition of $r$.
    \begin{align*}
    h_1(x_1) - h_1(x_2) & =  Q_q(x_1) - Q_r(x_2) \\
    & \leq  Q_q(x_1) - Q_q(x_2)  \mbox{~~ by $Q_q(x_2)\leq Q_r(x_2)$}\\
    & \leq  L \|x_1-x_2\| \mbox{~~~for $Q_q(x)$ is $L$-Lipschitz}
    \end{align*}
    Also, $h_1(x_1) - h_1(x_2)\geq 0$.
    Thus, we have that
    $$|h_1(x_1)-h_1(x_2)|\leq L \|x_1-x_2\|$$
    proving that $h_1(x)$ is Lipschitz continuous with parameter $L$.
    \end{proof}
        
    Let's now return to proving Claim~\ref{claim:2}. From \eqref{eq_d} and the definition of $k$, we have, for any subrectangle $j$,
    \begin{align}
    g_{f+1}(C_j) - L d_j & \geq  (1-\epsilon)\, g_{f+1}(C_k)  \nonumber \\
    \Rightarrow ~~ g_{f+1}(C_k) & \leq  \frac{1}{1-\epsilon} \left(g_{f+1}(C_j) - L d_j\right),  \label{e_hf}
    \end{align}
    Also, for any subrectangle $j$
    \begin{align}
    g_{f+1}(C_j) & \leq  h_1(C_j) \mbox{~~~~ by the definition of $g_{f+1}$ and $h_1$} \nonumber \\
    & \leq  \min_{x\in S_j} h_1(x) + L d_j \mbox{~~~~ due to Lemma \eqref{lemma:Lipschitz}} \label{e_hf2}
    \end{align}
    From \eqref{e_hf} and \eqref{e_hf2}, we have, for any $j$,
    \begin{align*}
    g_{f+1}(C_k) & \leq   \frac{1}{1-\epsilon} \left( 
    \left[ \min_{x\in S_j} h_1(x) + L d_j \right] - L d_j\right) \\
    & \leq  \left(\frac{1}{1-\epsilon}\right) \min_{x\in S_j} h_1(x)
    \end{align*}
    Since the above inequality holds for all $j$, and $\cup_j\, S_j = X$, we have
    $$g_{f+1}(C_k) \leq \left(\frac{1}{1-\epsilon}\right) \min_{x\in X} h_1(x).$$
    Now, because $h_{f+1}(\widehat{x}) \leq g_{f+1}(\widehat{x})=g_{f+1}(C_k)$, the above inequality implies that
    $$h_{f+1}(\widehat{x}) \leq \left(\frac{1}{1-\epsilon}\right) \min_{x\in X} h_1(x).$$
    This implies that at least $|\H|-f$ of the non-faulty cost functions are upper bounded by $\left(\frac{1}{1-\epsilon}\right) \, \min_{x\in X} h_1(x)$.
\end{proof}

Thus, the algorithm in this section results in
an increase in the upper bound by a factor of $\frac{1}{1-\epsilon}$.
As a trade-off, the algorithm in this section does not look for an exact global minimum of $g_{f+1}(x)$, and instead only approximates the minimum. The algorithm in Section \ref{sec:algo_1} finds the exact minimum of $g_{f+1}(x)$.

It is worth noting that the approximate algorithm above only evaluates the cost functions $Q_i(x)$'s at a finite number of locations. Therefore, such an algorithm can be extended to a distributed version under a server-based setting, where each agent $i$ obtains one cost function $Q_i(x)$, and sends the requested function values to the server, and the server conducts the procedure of finding $\widehat{x}$, without requiring sending of the whole cost functions. The algorithm can also be improved to determine $\widehat{x}$ more efficiently, similar to the DIRECT algorithm~\citep{jones2021direct} mentioned earlier in this section.
\section{Difficulty in using gradient descent methods}
\label{sec:gradient-method}

Gradient descent is a widely used and well-studied method for optimization problems \citep{boyd2004convex}. However, it might not be as trivial to apply gradient descent to the Byzantine min-max problem.
Consider this most direct approach: repeatedly update the current estimate $x_t$ using the gradient of a function with the $(f+1)$-largest function value at $x_t$ (breaking ties by randomly choosing), until convergence. Denote the function for the $t$-th update as $Q_{k_t}(x)$, we have
\begin{align}
    x_{t+1} = x_t - \eta_t \nabla Q_{k_t}(x_t),
\end{align}
where $\eta_t$ is the step size.

Recall \eqref{eqn:hf}. The above algorithm minimizes the function $g_{f+1}(x)$ using gradient descent. Yet gradient descent is not guaranteed to find a global minimum for non-convex functions, so it would not achieve the bound we claimed in the algorithm from Section~\ref{sec:algo_1}. In fact, the function $g_{f+1}(x)$ can be non-convex even without Byzantine agents. Furthermore, a Byzantine agent can manipulate the converged output of the gradient descent method to be arbitrarily far away from the true min-max solution, and the corresponding function value arbitrarily large, if an ``unlucky'' initial estimate $x_0$ is selected.
In other words, the output of the proposed gradient-based algorithm cannot be bounded at all.

The importance of this difficulty is that plain gradient descent can also be easily extended to scenarios of distributed optimization \citep{nedic2009distributed}, or Byzantine distributed optimization \citep{liu2021survey} specifically, when optimizing for the sum or average of a group of cost functions.
However, the discussion in this section indicates that we cannot easily apply gradient descent to a distributed Byzantine min-max problem \eqref{eqn:min-max}, and it would require other methods for the distributed setting.

\section{DIRECT-based distributed algorithm}
\label{sec:algo-distributed}

Consider the following distributed setting of the fault-tolerant min-max optimization problem, which we briefly discussed in Section~\ref{sec:intro}. In a system of 1 trusted server and $n$ agents, where each agent possesses one local cost function $Q_i(x)$. The communications are made between the server and each of the agents only. Among the $n$ agents, there are up to $f$ being Byzantine faulty, meaning they may behave arbitrarily. The ideal goal is to compute \eqref{eqn:min-max}. In the following discussions, we will continue using the notations $g_k(x)$ and $h_k(x)$ for all $k$ defined in the previous sections.

We now propose a distributed algorithm that is based on DIRECT \citep{jones1993lipschitzian, jones2021direct}, to solve fault-tolerant min-max problem on a pre-defined hypercube $X = \{x|u_i\leq x_i\leq v_i\}\in\R^d$ where $x_i$ is the $i$-th element of a vector $x$. Specifically, consider the following algorithm:

    \begin{algorithm}[h]
        \SetAlgoLined
        \caption{DIRECT-based distributed min-max algorithm}
        \label{alg}
        \textbf{Input:} $n$, $f$. A hypercubical region $X$. Each agent $i$ has its cost function $Q_i(x)$.
        
        \vspace{2pt}
    
        For initialization, normalize $X$ to a unit hypercube of the same dimension $d$. Let $c_1$ be the center point of this hypercube and evaluate $Q_i(c_1)$ for all $i$. Set $g^{\mathrm{out}}_{f+1} = \underset{i\in[n]}{\mathrm{rank}_{f+1}}Q_i(c_1)$, $m=1$, and $t=0$.
        
        On each iteration $t$,
        \begin{enumerate}[nosep, label=\textbf{Step \arabic*:}]
            \item Identify the set $S$ of potentially optimal (hyper)-rectangles.
            \item Select any rectangle $j\in S$.
            \item Decide where to sample within $j$ and how to divide the rectangle into subrectangles. Request $Q_i(x)$ from all sample points.
            \item Update $g^{\mathrm{out}}_{f+1} = \min\{g^{\mathrm{out}}_{f+1}, \underset{i\in[n]}{\mathrm{rank}_{f+1}}Q_i(x)\}$ for all $x$ sampled in this iteration, and set $m = m+\Delta m$, where $\Delta m$ is the number of new points added.
            \item Set $S = S-\{j\}$. Repeat from Step 2 until $S=\varnothing$.
            \item Repeat from Step 1 until reaching the limit of iteration. The algorithm then output $g_{f+1}^{out}$ and its corresponding sampled point $\widehat{x}$.
        \end{enumerate}
    \end{algorithm}
    
    A rectangle $i$ is said to be \textit{potentially optimal} if its center $c_i$ and center-vertex distance $d_i$ satisfy the following \citep{jones1993lipschitzian}: there exists some $K$ such that
    \begin{align}
        g_{f+1}(c_i) - Kd_i &\leq g_{f+1}(c_j) - Kd_j, ~~~\forall j, \textrm{ and} \label{eqn:cond-1}\\
        g_{f+1}(c_i) - Kd_i &\leq g^{\mathrm{out}}_{f+1} - \epsilon\mnorm{g^{\mathrm{out}}_{f+1}}. \label{eqn:cond-2}
    \end{align}

    Algorithm~\ref{alg} essentially optimizes the function $g_{f+1}(x)=\underset{i\in[n]}{\mathrm{rank}_{f+1}Q_i(x)}$, the same as what we did in the previous sections. In Step 3 of Algorithm~\ref{alg}, the division is performed according to the following subroutine \citep{jones2021direct} specified in Algorithm~\ref{alg-divide}. Intuitively, the algorithm divides each side of a potentially optimal rectangle into 3, in an order such that one of the largest divided subrectangles would have the smallest function value at the center.
    \begin{algorithm}
        \caption{Procedure for dividing rectangles \citep{jones1993lipschitzian}}
        \label{alg-divide}
        \textbf{Input:} a potentially optimal rectangle $S$
        \begin{enumerate}[label=\textbf{S\arabic*:}, nosep]
            \item Identify the set $M$ of dimensions with the maximum side length of the rectangle. Let $\delta$ be $1/3$ of this maximum side length.
            \item Sample the function at points $\boldsymbol{c}\pm\delta\boldsymbol{e}_k$ for all $k\in M$, where $\boldsymbol{c}$ is the center of $S$, and $\boldsymbol{e}_k$ is the $k$-th unit vector.
            \item Divide the rectangle $S$ into thirds along the dimensions in $M$, starting with the dimension with the lowest value of $w_k=\min\{\underset{i\in[n]}{\mathrm{rank}_{f+1}}Q_i(\boldsymbol{c}+\delta\boldsymbol{e}_k), \underset{i\in[n]}{\mathrm{rank}_{f+1}}Q_i(\boldsymbol{c}-\delta\boldsymbol{e}_k)\}$ (breaking ties arbitrarily), and continuing to the dimension with the highest $w_k$.
        \end{enumerate}
    \end{algorithm}

    To discuss the performance of the algorithm, we will use the following claim.
    \begin{claim}
        A piecewise function of finite pieces of $L$-Lipschitz functions over a compact convex set $D$ is also $L$-Lipschitz.
        \label{lemma:piecewise-lipschitz}
    \end{claim}

    \begin{proof}
        Consider any two points $a,b$ in $D$ where the piecewise function $f$ is defined. Suppose that each piece $i$ of $f$ is defined over a closed region $D_i\subset D$ such that $D = \bigcup_i D_i$ and for all $i\neq j$, regions $D_i$ and $D_j$ may only overlap on their boundaries.
        
        Since $D$ is convex, every point on the line between $a$ and $b$ is in $D$. The line $ab$ crosses boundaries of $D_i$'s for a finite times. Let those crossing points be $\{x_1,...,x_k\}$ such that, when also letting $x_0=a$ and $x_{k+1}=b$, $x_i$'s are all on the boundaries of two neighboring pieces of $f$, and $x_i$ and $x_{i+1}$ are always on the same piece. Also,
        \begin{align}
            \norm{x_0-x_{k+1}} = \norm{x_0-x_1} + ... + \norm{x_k-x_{k+1}}.
        \end{align}

        By the Lipschitz condition, there exists $L>0$, such that 
        \begin{align}
            \mnorm{f(x_i)-f(x_{i+1})} \leq L\norm{x_i-x_{i+1}}.
        \end{align}
        By triangle inequality,
        \begin{align}
            \mnorm{f(x_0)-f(x_{n+1})} &= \mnorm{f(x_0) - f(x_1) + f(x_1) - f(x_2) + ... + f(x_k) - f(x_{k+1})} \nonumber \\
            &\leq \mnorm{f(x_0) - f(x_1)} + \mnorm{f(x_1) - f(x_2)} + ... + \mnorm{f(x_k) - f(x_{k+1})} \nonumber \\
            &\leq L\norm{x_0-x_1} + L\norm{x_1-x_2} + ... + L\norm{x_n-x_{n+1}} \nonumber \\
            &\leq L\norm{x_0-x_{n+1}},
        \end{align}
        therefore, the function $f$ is also $L$-Lipschitz.
    \end{proof}

    \subsection{Convergence}
    In the following two claims, we examine the convergence and accuracy of the algorithm. Note that we do not need to check the Lipschitzness of the cost functions, 
    and therefore, $g_{f+1}$ can be non-Lipschitz.

    \begin{claim}[cf. \citep{jones1993lipschitzian}]
        Suppose non-faulty cost functions are $L$-Lipschitz after normalizing $X$. Also, suppose cost functions from all agents are non-negative. The algorithm produces a set of sampled points is everywhere dense in the hypercube. That is, for any point $x$ in the hypercube and $\delta>0$, the algorithm will eventually sample a point within $\delta$ distance of $x$.
        \label{claim:min-max-everywhere-dense}
    \end{claim}
    This claim and the following proof largely follow \citep[Section 5]{jones1993lipschitzian}. We write the proof in full for completeness.
    \begin{proof}
        Recall that every time a subrectangle is divided, it is divided by 3 on its largest side. Therefore, for any subrectangle, the only possible side lengths are $3^{-k}$ for $k=0,1,2,...$. Also, for any subrectangle, no side of length $3^{-(k+1)}$ can be divided, until all sides of length $3^{-k}$ have been divided. Therefore, after $r$ divisions, the subrectangle will have $q=r-kd$ sides of length $3^{-(k+1)}$ and $p=d-q$ sides of length $3^{-k}$, where $k=\lfloor r/d\rfloor$ and $d$ is the dimension of space the cost functions are defined on. We can then write down the distance from the center of the subrectangle to its vertices as follows:
        \begin{align}
            \widetilde{d} = \sqrt{p\cdot(3^{-k})^2 + q\cdot(3^{-(k+1)})^2}/2.
        \end{align}
        It is obvious, that as $r\rightarrow\infty$, we have $k\rightarrow\infty$ and therefore, the center-vertex distance of any subrectangle $\widetilde{d}\rightarrow0$.

        Suppose $r_t$ is the fewest number of divisions undergone by any subrectangle when the algorithm enters iteration $t$. We claim that $\lim_{t\rightarrow\infty}r_t=\infty$. Show by contradiction: Suppose $\lim_{t\rightarrow\infty}r_t\neq\infty$. That is, there exists some $t'$, such that after iteration $t'$, $r_t$ stops increasing. In other words, $\lim_{t\rightarrow\infty}r_t=r_{t'}$. By definition, at the end of iteration $t'$, there will be a finite number of subrectangles which have been divided exactly $r_{t'}$ times. Let that number be $N$. All these subrectangles have the same and the largest center-vertex distance, but they have different center points and therefore may have different function values of $g_{f+1}$ at their center points. Let rectangle $j$ be the one with the best function value at the center point. In the next iteration, rectangle $j$ will be potentially optimal, if the two conditions in \eqref{eqn:cond-2} can be satisfied. Consider $\widetilde{L}>\max\{L_1,L_2\}$, where
        \begin{align}
            L_1 &= \frac{g_{f+1}(c_j)-g_{f+1}^{\textrm{out}}+\epsilon\mnorm{g_{f+1}^{\textrm{out}}}}{d_j}, \label{eqn:def-l1} \\
            L_2 &= \max_{i\textrm{ s.t. }d_i<d_j}\frac{g_{f+1}(c_j)-g_{f+1}(c_i)}{d_j-d_i}.
            \label{eqn:def-l2}
        \end{align}
        
        Consider \eqref{eqn:def-l1}. Let $x^*_1\in\arg\min h_1(x)$. 
        By Claim~\ref{lemma:piecewise-lipschitz}, since $h_1(x)$ is a piecewise function of non-faulty cost functions, and non-faulty cost functions are all $L$-Lipschitz, $h_1(x)$ is also $L$-Lipschitz. Therefore, we have
        \begin{align}
            h_1(c_j)-\min h_1(x) \leq L\norm{c_j-x^*_1}.
        \end{align}
        $X\subset\R^d$ is a $d$-dimensional unit hypercube, so $\norm{c_j-x^*_1}\leq\sqrt{d}$. Thus, since $g_{f+1}(x)\leq h_1(x)$ by definition, we have
        \begin{align}
            g_{f+1}(c_j) \leq h_1(c_j) \leq L\sqrt{d} + \min h_1(x) <\infty
            \label{eqn:l_1_bound}
        \end{align}
        being finite.
        Therefore, since $d_j$ is a finite value and $g_{f+1}^{out}\leq g_{f+1}(c_j)<\infty$, $L_1$ is also finite.
        
        Now consider \eqref{eqn:def-l2}. For any subrectangle $i$ at the end of iteration $t'$, since $h_{f+1}(x)\leq g_{f+1}(x)\leq h_1(x)$, 
        \begin{align}
            g_{f+1}(c_j)-g_{f+1}(c_i) &\leq h_1(c_j) - h_{f+1}(c_i) \nonumber \\
            &= h_1(c_j) - h_{f+1}(c_j) + h_{f+1}(c_j) - h_{f+1}(c_i)
            \label{eqn:l2-finite-1}
        \end{align}
        By Claim~\ref{lemma:piecewise-lipschitz}, since $h_{f+1}(x)$ is a piecewise function of non-faulty cost functions, and non-faulty cost functions are all $L$-Lipschitz, $h_{f+1}(x)$ is also $L$-Lipschitz. Also, since $c_i,c_j\in X$ where $X\subset\R^d$ is a $d$-dimensional unit hypercube, we have $\norm{c_i-c_j}\leq \sqrt{d}$. Therefore,
        \begin{align}
            h_{f+1}(c_j)-h_{f+1}(c_i) \leq L\norm{c_j-c_i}\leq L\sqrt{d}.
            \label{eqn:l2-finite-2}
        \end{align}
        Therefore, 
        \begin{align}
            g_{f+1}(c_j)-g_{f+1}(c_i) &\leq h_1(c_j) - h_{f+1}(c_j) + h_{f+1}(c_j) - h_{f+1}(c_i)  \tag{by \eqref{eqn:l2-finite-1}}\nonumber \\
            &\leq h_1(c_j) + \left(h_{f+1}(c_j)-h_{f+1}(c_i)\right) \nonumber \tag{cost functions are non-negative} \\
            &\leq h_1(c_j) + L\sqrt{d} \nonumber \tag{by \eqref{eqn:l2-finite-2}} \\
            &\leq 2L\sqrt{d} + \min h_1(x) < \infty, \tag{by \eqref{eqn:l_1_bound}}
        \end{align}
        and $L_2$ is also finite.
        
        \comment{Nitin: what if there are multiple rectangles of largest diameter? Which of these is j?\\\textcolor{black}{shuo: by (17) only c\_i with the smallest g\_f+1(c\_i) will be selected, if d\_i's are tied}}
        Therefore, it is always possible to choose a finite $\widetilde{L}$, such that \eqref{eqn:cond-2} are satisfied, meaning the subrectangle $j$ will be divided in iteration $t'+1$ (as well as other rectangles of the same diameter as $d_j$ and the same center value as $g_{f+1}(d_j)$), and no longer has $r_{t'}$ divisions. In iteration $t'+2$, following the same argument, the subrectangle with $r_{t'}$ divisions and the best function value at the center point will be divided. ... Until in iteration $t'+N$, all subrectangles that have $r_{t'}$ divisions will be divided, and $r_{t'+N}< r_{t'}$. This is contradictory to our assumption that $\lim_{t\rightarrow\infty}r_t=r_{t'}$. Hence, it follows that the maximum center-vertex distance must approach 0 when $t\rightarrow\infty$.

        Thus, given any $\delta>0$, there exists some $T$, such that after $t>T$ iterations, every subrectangle has center-vertex distance smaller than $\delta$, implying that the sampled points of the algorithm are everywhere dense. Hence the proof.
    \end{proof}

    \begin{claim}
        \label{claim:min-max-bound}
        Suppose non-faulty cost functions are $L$-Lipschitz after normalizing $X$. Also, suppose cost functions from all agents are non-negative. Algorithm~\ref{alg} guarantees its output $c_j$ is bounded as
        \begin{align}
            \min_x h_{f+1}(x) \leq g_{f+1}(c_j) \leq \min_x h_1(x)+L\cdot\max_id_i.
        \end{align}
    \end{claim}
    \begin{proof}
         Any Byzantine agents may send function values that do not satisfy any Lipschitzness. For analysis purposes, we denote $l$ to be a rectangle with center $c_l$, such that at the time the algorithm stops, $l=\arg\min_i\left[g_{f+1}(c_i)-L\cdot d_i\right]$.
    
        Note that without checking Lipschitzness, $g_{f+1}$ can be non-Lipschitz due to the non-Lipshictz pieces sent by faulty agents. For $l$, we have 
        \begin{align}
            g_{f+1}(c_l)-L\cdot d_l\leq\min_i\left[g_{f+1}(c_i)-L\cdot d_i\right].
        \end{align}
        That is, for all rectangle $i$ at the time the algorithm terminates,
        \begin{align}
            g_{f+1}(c_l)-L\cdot d_l\leq g_{f+1}(c_i)-L\cdot d_i.
            \label{eqn:upperbound-cl}
        \end{align}
        Note that for all $i$, by the definitions of $g_{f+1}(x)$ and $h_1(x)$,
        \begin{align}
            g_{f+1}(c_i)-L\cdot d_i \leq h_1(c_i) - L\cdot d_i.
        \end{align}
        Since $h_1(x)$ is $L$-Lipschitz, there exists an $k$ such that
        \begin{align}
            h_1(c_k) - L\cdot d_k \leq \min_x h_1(x).
        \end{align}
        Therefore, 
        \begin{align}
            g_{f+1}(c_k)-L\cdot d_k\leq \min_xh_1(x).
            \label{eqn:upperbound-ck}
        \end{align}
        Combining \eqref{eqn:upperbound-cl} and \eqref{eqn:upperbound-ck}, we have
        \begin{align}
            &g_{f+1}(c_l)-L\cdot d_l \leq \min_x h_1(x).
        \end{align}
        Or,
        \begin{align}
            &g_{f+1}(c_l)\leq \min_x h_1(x) + L\cdot d_l\leq \min_x h_1(x) +L\cdot\max_id_i.
        \end{align}
        Now consider the output of the algorithm $c_j$. By the algorithm, 
        \begin{align}
            g_{f+1}(c_j)=\min_ig_{f+1}(c_i)\leq g_{f+1}(c_l).
        \end{align}
        Combining the two results above, we have
        \begin{align}
            g_{f+1}(c_j)\leq\min_xh_1(x)+L\cdot\max_id_i.
        \end{align}
        On the other hand, by the definitions of $g_{f+1}(x)$ and $h_{f+1}(x)$,
        \begin{align}
            g_{f+1}(c_j) \geq \min_x g_{f+1}(x)\geq \min_x h_{f+1}(x).
        \end{align}
        Therefore, 
        \begin{align}
            \min_x h_{f+1}(x) \leq g_{f+1}(c_j) \leq \min_x h_1(x)+L\cdot\max_id_i.
        \end{align}
    \end{proof}
    That is, the output of the algorithm is lower bounded by the minimum of the rank $f+1$ non-faulty function, and upper bounded by the minimum of the rank $1$ non-faulty function, plus a constant that is proportional to the size of the largest subrectangle when the algorithm stops.

    Combining Claims~\ref{claim:min-max-everywhere-dense} and \ref{claim:min-max-bound}, the algorithm is known to always divide its largest subrectangles in any iteration, therefore $\lim_{t\rightarrow\infty}\max_id_i\rightarrow0$, and the output of the algorithm is asymptotically bounded by
    \begin{align}
        \min_x h_{f+1}(x) \leq g_{f+1}(c_j) \leq \min_x h_1(x).
    \end{align}

\section{Summary}

In this paper, we considered a new problem of Byzantine fault-tolerant min-max optimization, its formulation and a possible direction of approaching it. We provided some tight bounds on the output of an algorithm for this problem. We also presented an approximate variation for solving this problem.
We discussed that it is difficult to apply gradient-based methods directly to this problem and its implication for solving the distributed version of this problem.
We then presented a distributed algorithm for Byzantine fault-tolerant min-max optimization and analyzed its convergence and performance.
This paper opens up a new line of research. Future work includes the development of more efficient algorithms and {\em distributed} solutions to Byzantine min-max optimization.




\bibliographystyle{plainnat}
\bibliography{bib}

\end{document}